\documentclass[twocolumn,showpacs,amsmath,amssymb, showkeys]{revtex4}

\usepackage{graphicx}% Include figure files
\usepackage{dcolumn}% Align table columns on decimal point
\usepackage{bm}% bold math

\begin{document}

\title{Testing gravity at the Second post-Newtonian level through
gravitational deflection of massive particles}
\author{A. Bhadra} 
\email{aru_bhadra@yahoo.com}
\affiliation{Admn. Block, University of North Bengal, Siliguri, WB 734013 
INDIA}
\author {K. Sarkar}
\author{K. K. Nandi} 
\email{kamalnandi1952@yahoo.com}
\affiliation{Department of Mathematics, University of North Bengal, Siliguri, WB 734013 
INDIA}

\begin{abstract}
Expression for second post-Newtonian level gravitational deflection angle of
massive particles is obtained in a model independent framework. Ambiguities
associated with massless particle deflection can be avoided by using massive
particles. Comparison of theoretical values with the observationally
constructed values of post-Newtonian parameters for massive particles offers
the future possibility of testing at that level competing gravitational
theories as well as the equivalence principle.
\end{abstract}

\maketitle 

\section{Introduction}

A consequence of general relativity (GR) is that light rays are deflected by
gravity. Historically, observations of this aspect of gravity provided the
first proof in favour of GR. The effect is nowadays routinely used as a tool
to study various features of the universe, such as viewing fainter or
distant sources, estimating the masses of galaxies etc. [1].

Like photons, particles having masses are also deflected by gravity. The
general relativistic corrections in the equation of motion of a massive test
particle moving in bound orbits have been studied with great accuracy in the
literature [2]. These studies have great relevance in comparing theory with
observations of gravitational waves from compact binary systems. However,
for {\it unbound} orbits of massive particle, the question of accuracy
beyond first order has not received sufficient attention as of now. The main
reason could be that the observational aspect of unbounded massive particles
was not very practical: there was no known astrophysical source of free
point particles that can be detected easily with good angular precision. The
situation seems to have improved somewhat. Recent theoretical studies [3]
favor the existence of local astrophysical sources of relativistic neutral
particles like neutrons and neutrinos with observable fluxes. Besides, high
energy neutrons are produced during solar flares [4]. Moreover, with the
advent of new technology new experiments have been proposed [5], primarily
to study gravitational deflection of light with high precision, in which
laser interferometry will be employed between two micro-spacecrafts whose
line of sight pass close to the sun. Hence there might be a possibility that
in the future, neutron or some other neutral particle may be used in a
similar experiment instead of photon thus providing an opportunity of
studying gravitational deflection of massive particles. Henceforth, we use
the abbreviation for post-Newtonian as PN such that first-PN effect is of
the order of $\left( 1/\rho \right) $, second-PN effect is of the order of $%
\left( 1/\rho ^{2}\right) $ and so on.

The expected angular precision of the planned astrometric missions using
optical interferometry is at the level of microarcseconds ($\mu$ arcsec)
and hence these experiments would measure the effects of gravity on light
at the second-PN order ($c^{-4}$). Though measurements with massive
particles at the level of microarcsecond accuracy is way beyond the present
technical capability, it can still be cautiously hoped that astrometric
missions in the distant future using massive particle interferometry would
have angular precisions close to that to be obtained using laser (optical)
interferometry. Whatever be the technical scenario, a study of theoretical
aspects of gravitational deflection of massive particles at the second-PN
approximation is useful in its own right. (To our knowledge, the deflection
angle for massive particles has been theoretically estimated in the
literature with an accuracy of only first-PN order [6] so far).

Accordingly, in the present article, the second-PN contribution to the
gravitational deflection of massive particles by a gravitating object will
be estimated in a model independent way but with a special emphasis on the
sun as gravitating object. The corresponding PN parameters for light
deflection then follow as a corollary. The importance of this investigation
is manifold: It allows us to obviate the ambiguities related with photon
deflection in that order; it allows us to construct the true coordinate
solar radius from measurements and moreover, it constitutes a possible
futher test of the equivalence principle. These are discussed at the end.

\section{Gravitational deflection of massive particles at second
post-Newtonian order}

We consider the general static and spherically symmetric spacetime in
isotropic coordinate which is given by (we use the geometrized units \textit{%
i.e.} $G=1$, $c=1$) 
\begin{equation}
ds^{2}=-B(\rho )dt^{2}+A(\rho )\left( d\rho ^{2}+\rho ^{2}d\theta ^{2}+\rho
^{2}sin^{2}\theta d\phi ^{2}\right)
\end{equation}%
Restricting to orbits in the equatorial plane ($\theta =\pi /2$), the
expression for the deflection angle for particles moving with a velocity $V$
as measured by an asymptotic rest observer can be written as [7] 
\begin{equation}
\alpha (\rho _{o})=I(\rho _{o})-\pi
\end{equation}%
with (see appendix) 
\begin{equation}
I(\rho _{o})=2\int_{\rho _{o}}^{\infty }\frac{d\rho }{\rho ^{2}}%
A^{-1/2}(\rho )\left[ \frac{1}{J^{2}}\left( B^{-1}(\rho )-E\right) -\frac{1}{%
\rho ^{2}A}\right] ^{-\frac{1}{2}}\;
\end{equation}%
where $\rho _{o}$ being the distance of the closest approach, 
\begin{equation}
J=\rho _{o}\left[ A(\rho_{o}) (B^{-1}(\rho _{o})-E ) \right]^{1/2}
\end{equation}%
and 
\begin{equation}
E=1-V^{2}
\end{equation}

The PN formalism in some orders [7,8] is usually employed to describe the
gravitational theories in the solar system and also to compare predictions
of GR with the results predicted by an alternative metric theory of gravity.
This method actually is an approximation for obtaining the dynamics of a
particle (in a weak gravitational field under the influence of a slowly
moving gravitational source) to one higher order in $\frac{M}{\rho }$ ($M$
is the mass of the static gravitating object) than given by the Newtonian
mechanics. Following the PN expansion method, we assume the metric tensor is
equal to the Minkowski tensor $\eta _{\mu \nu }$ plus corrections in the
form of expansions in powers of $\frac{M}{\rho }$ and considering up to the
second-PN correction terms, we have 
\begin{equation}
B(\rho )=1-2\frac{M}{\rho }+2\beta _{i}\frac{M^{2}}{\rho ^{2}}-\frac{3}{2}%
\epsilon _{i}\frac{M^{3}}{\rho ^{3}}
\end{equation}%
\begin{equation}
A(\rho )=1+2\gamma _{i}\frac{M}{\rho }+\frac{3}{2}\delta _{i}\frac{M^{2}}{%
\rho ^{2}}
\end{equation}%
$\beta _{i},\gamma _{i}$ are the PN parameters (also known as the Eddington
parameters), $\delta _{i}$ and $\epsilon _{i}$ can be considered as the
second-PN parameters, $i$ stands for either $%
\gamma $ or $m$ denoting photons or massive particles respectively. Several
of these parameters are different for different theories. In GR, all of them
are equal to $1$ as can be readily checked by expanding the Schwarzschild
metric.  

We should note that the metric coefficients above are independent of any
specific model; they result solely from the assumption of central symmetry.
Starting with the expansion (6) and (7) \textit{per se}, the expression for
the angle of deflection for unbound particles up to the second-PN order
follows from Eq.(3) and when $\frac{2M}{\rho }<<V^{2}$, it works out to 
\begin{equation}
\alpha _{m}=a_{m}\frac{M}{\rho _{o}}+b_{m}\left( \frac{M}{\rho _{o}}\right)
^{2}
\end{equation}%
where 
\begin{equation}
a_{m}=2\left(\gamma _{m}+ \frac{1}{V^{2}}\right)
\end{equation}%
\begin{equation}
b_{m}=\frac{3\delta _{m}\pi }{4}+\left(2+2\gamma _{m}-\beta
_{m}\right)\frac{\pi}{V^{2}} -2\left( \gamma _{m}+\frac{1}{V^{2}}\right)^{2} 
\end{equation}%
The above expressions are also valid for massless particles ($\alpha
_{\gamma }$, $a_{\gamma }$, $b_{\gamma }$) as may be seen under the
substitution $V=1$. Clearly the deflection angle would be larger for
particles in comparison to that of photons. Our calculation shows that the
term representing the second order effect ($b_{m}$) contains only the three
parameters $\beta _{m}$, $\gamma _{m}$, $\delta _{m}$ and does not contain $%
\epsilon _{m}$, a cubic order contribution. 
This implies that calculation of the deflection of \textit{unbound} particle orbits (including
photons) by gravity to any given order needs only the knowledge of every
term to that order in the expansions. In other words, to second-PN order, one
needs to consider both in $\ g_{oo\text{ }}$ and $g_{ij}$ terms only up to $%
\frac{M^{2}}{\rho ^{2}}.$ Similarly, to third-PN order, which is not our interest here, we would need expansions of both the metric components up to order $\frac{M^{3}}{\rho ^{3}}
$ and so on. This is in contrast to the case of planetary 
dynamics (\textit{bound} orbits) where the calculation typically requires knowledge of $g_{oo}$
more accurately than $g_{ij}$ (For instance, to calculate the planetary
precession to the order of $M$, one expands $g_{oo}$ up to $\frac{2\beta
_{m}M^{2}}{\rho ^{2}}$ while $g_{ij}$ is expanded up to only $\frac{2\gamma
_{m}M}{\rho }$; for next order accuracy, one would need to consider the
complete expansion as given in Eqs.(6), (7) above so that the parameters $%
\delta _{m}$, $\epsilon _{m}$ become important in this case). The deflection angle $\alpha
_{m} $ for the Schwarzschild spacetime can be obtained by taking $\beta
_{m}=\gamma _{m}=\delta _{m}=1$.

The deflection angle also can be expressed in terms of coordinate
independent variables, such as the impact parameter $b$ which is the
perpendicular distance from the center of the gravitating object to the
tangent to the geodesic at the closest approach. In that case, $\rho $ has
to be replaced by $b$ in Eq.(8), the Eq.(9) would remain unaltered but
Eq.(10) would change to 
\begin{equation}
b_{m}=\frac{3\delta _{m}\pi }{4}+ \left(2+2\gamma _{m}-\beta
_{m} \right)\frac{\pi}{V^{2}} + 2\left( \gamma _{m} + \frac{1}{V^{2}}\right) \left( 1-\frac{1}{V^{2}}\right) %
\end{equation}%
Since impact parameter is the ratio of the angular momentum and energy of
the particle as measured by an observer at rest far from the gravitating
object, it is a formally measurable quantity but is not very suitable for
practical measurements [9].

\section{Other significant effects}

In the present work, the mass distribution of the gravitating object is
assumed to be mainly spherically symmetric; any deviation from such symmetry
would produce their effects. The effect of quadrupole moment of the mass
distribution on the deflection angle is proportional to $\frac{J_{Q}MR^{2}}{%
\rho _{o}^{3}}$, where $R$ is the average radius of sun. Thus, even a small
quadrupole moment parameter $J_{Q}$ could produce significant contribution
to deflection. However, the effect is limited largely to the first-PN order (%
$\sim 0.1$ $\mu$ arcsec) while in the second-PN order the effect is
too small ($\sim 10^{-7}-10^{-8}$ $\mu$ arcsec). If the gravitating
object also has angular momentum, its effect on the deflection angle
contributes to the second-PN order but it can be separated out. All these
are discussed below.

\subsection{Effect of quadrupole moment of the mass distribution}

Theoretical value of solar quadrupole moment $J_{Q}$, though it depends
strongly on solar model used, is very small, of the order of $10^{-7}$
[5,8]. Since our study is aimed at sun as the gravitating object we have
ignored higher order terms involving $J_{Q}$. Thus due to the quadrupole moment of
the mass distribution the effective mass parameter becomes $M_{eff}=M\left[
1+\frac{J_{Q}R^{2}}{2\rho ^{2}}\left( 3cos^{2}\theta -1\right) \right] $
which leads to the following corrections in the components of the metric
tensors [5,8,10]: 
\begin{equation}
\delta g_{oo}(\rho )=J_{Q}\frac{MR^{2}}{\rho ^{3}}\left( 3cos^{2}\theta
-1\right)
\end{equation}%
and 
\begin{equation}
\delta g_{jk}(\rho )=-\delta _{jk}\gamma _{i}J_{Q}\frac{MR^{2}}{\rho ^{3}}%
\left( 3cos^{2}\theta -1\right)
\end{equation}%
where $R$ is the average radius of the mass distribution and $\theta $ is
the angle between radius vector and the $z$-axis and hence in the equatorial
plane $\theta =\pi /2$. In the equatorial plane, the deflection caused by
the quadrupole moment calculates to 
\begin{equation}
\alpha _{QM}=\frac{2J_{Q}MR^{2}}{\rho _{o}^{3}}\left( \frac{\gamma _{m}}{3}+%
\frac{1}{V^{2}}\right)
\end{equation}

\bigskip Assuming $R\sim \rho _{o}$ at the closest approach to the sun and
taking $V=0.75$, $\beta _{m}=\gamma _{m}=\delta _{m}\sim 1$, for sun $\frac{%
M_{\odot }}{R_{\odot }}=2.12\times 10^{-6}$, this first-PN quadrupole term $%
\alpha _{QM}$ $\sim 0.1$ $\mu$ arcsec. It is roughly $7$ orders of
magnitude less than the first-PN deflection $a_{m}\frac{M}{\rho _{o}}\sim 0.8
$ $\sec $ and is more than one order of magnitude less than
the second-PN contribution $b_{m}\left( \frac{M}{\rho _{o}}\right) ^{2}\sim
3.4$ $\mu$ arcsec. We have not displayed the next higher order
quadrupole terms involving $J_{Q}^{2}$ and second order terms in $1/\rho ^{2}
$ containing $J_{Q}$ here because they have magnitude in the range $10^{-7}$
to $10^{-8}$ $\mu$ arcsec, too small to be of any practical
significance. We can justifiably ignore these second-PN quadrupole
contribution. The quadrupole contribution to the deflection of light is
given in Ref.[11].

\subsection{Effect of rotation}

The angular momentum of the gravitating object is assumed small as in the
case of sun. The resulting leading term of the relevant metric tensor is 
\begin{equation}
g_{oi}=\frac{4Ma}{\rho }
\end{equation}%
where $a$ is the angular momentum per unit mass of the object. The
contribution of the rotation to the deflection angle is then given by [6] 
\begin{equation}
\alpha _{rot}=\frac{4MaV}{\rho _{o}^{2}}
\end{equation}%
The value of $a$ can be positive or negative depending on the direction of
rotation. When the angular momentum of the gravitating object is
antiparallel with the direction of the incoming particle, $a$ is positive
and hence rotation causes larger deflection whereas for parallel angular
momentum, $a$ is negative and the deflection angle will be less. Thus the
rotational effect can be easily separated out from other contributions by
studying the deflection of particles at two opposite sides of the
gravitating object.

The gravitational deflection angle of light with an accuracy up to second-PN
order readily follows from Eqs.(8)-(10), (14) and (16) using $V=1$.

\section{Discussion}

The study of gravitational deflection of massive particles is important for
several reasons which are discussed below.

First of all, observations of gravitational deflection of massive particles
with $\mu$ arcsec precision could probe the gravitational theories at
the second-PN level without any ambiguity. The second order predictions of
gravitational deflection of light as evolved from different studies are
found to be ambiguous. This is because of erroneous identification of
theoretical deflection variables with the observables or measured
quantities. For instance, when light ray just grazes the limb of the Sun and
the isotropic radial coordinate $\rho _{o}$ is identified with the measured
(under \textit{Euclidean } approximation) solar radius, it gives a second-PN
contribution of $\sim 3.5$ $\mu$ arcsec to deflection angle in GR
[12]. But if one uses the standard Schwarzschild coordinates instead, the
second-PN contribution to the deflection angle of light in GR following from
Eqs.(8)-(10) would be $\left[ \frac{15\pi }{16}-1\right] \frac{4M^{2}}{%
r_{o}^{2}}$ which is numerically about $7\;\mu$ arcsec for light ray
grazing the limb of the sun, provided the distance of closest approach in
such coordinates is identified with the measured radius of the sun. The
deflection angle can also be expressed in terms of coordinate independent
variables, such as the impact parameter $b$. In that case, the second-PN
contribution to deflection angle in GR becomes $\frac{15\pi }{16}\frac{4M^{2}%
}{b^{2}}$ and when at closest approach, $b$ is identified as the measured
solar radius for light grazing the sun, the magnitude of second-PN
deflection angle is $\sim 11\;\mu$ arcsec [13]. Thus there exists a
great confusion about the prediction of GR (or in fact of any viable
gravitational theory) at the second-PN order.

The fundamental reason for such an ambiguity is that the measurements of
solar radius usually employ Euclidean geometry as an approximation [6]
whereas the angle of gravitational deflection or other GR effects are
principally based on the consideration of curved spacetime. But, comparing
points in two different geometries i.e., in curved spacetime and flat
spacetime is totally meaningless [14]. The Euclidean approximation works
tolerably well only up to the first order, that is, in weak field gravity
caused by a source like Sun. The magnitude of the second order contribution
is, however, of the same order as the error that arises due to such an
approximation. Hence the numerical value of gravitational deflection angle
of light can not be unambiguously predicted at the level of second-PN order
within the theoretical scheme in vogue. Since the deflection angle for
massive particles depends also on the \textit{velocity} of the particle, the
stated ambiguity can be easily avoided by measuring deflection angles for
two or more velocities of the probing massive particle.

In GR, a coordinate length like $\rho _{o}$ is not directly measurable, it
can only can be indirectly ``constructed'' from the values of actual
measurements. The PN parameters and also the otherwise unknown coordinate
solar radius $\rho _{o}$ (or equivalently, $r_{o}$ in standard Schwarzschild
coordinates) can be constructed through least square fitting with the
measured deflection angles $\alpha _{m}$ and probing velocities $V$ using
the Eqs. (8)-(10). The idea is that the values of coordinates, $\rho _{o}$
and $r_{o}$, which refer to the same radial point, should be treated more
like other PN parameters ($\beta _{m}$, $\gamma _{m}$, $\delta _{m}$) due to
the fact that the ``flat geometry'' spacetime points can not
be algebraically identified in a curved spactime [14]. Technically, however,
the flat radial distances can be constructed by using metric gravity itself
(Eddington expansion) in terms of a large set of unknown PN parameters ($%
\beta _{m}$, $\gamma _{m}$, $\delta _{m}$) including $\rho _{o}$ by fitting
them with the observed data [7]. This method has been adopted, for example,
by Shapiro and his group in the radar echo delay observations [15, 16]. The
resulting parameter values can then be compared with the theoretical
predictions of deflection in GR as well as in other competing theories (like
Brans-Dicke theory) in the second-PN order involving both massive and
massless partcles.

The study of gravitational deflection of massive particles is also important
in the context of testing the weak equivalence principle which is one of the
fundamental postulates of general relativity. The principle states that the
trajectory of a freely falling object is independent of its internal
structure and composition. In other words all particles are coupled with
spacetime geometry universally. The principle has been tested with great
accuracy through different experiments, notable among them are the E\"{o}tv%
\"{o}s type experiments [10] where comparison of gravitational and inertial
masses of objects are made by measuring their acceleration in a known
gravitational field. For massless particles like photons, however, such
measurements obviously can not be performed. Instead, in such situation the
principle is tested by examining whether the gravitational (second-PN)
coupling parameter $\gamma $ is universal for all particles, massive or
massless. On the basis of supernova 1987 neutrino and optical data [17], a
limit of $\mid \gamma _{\gamma }-\gamma _{m}\mid \leq 3.4\times 10^{-3}$ has
actually been found [18]. However, the mass of a neutrino $m_{\nu _{e}}$ is
very small (if not zero); the present upper limit being $m_{\nu _{e}}\leq 3$
eV. Hence, a more conclusive experiment would be to examine whether the
gravitational couplings for photon and massive particles (other than
neutrinos) are same or not. The observational value of $\mid \gamma _{\gamma
}-\gamma _{m}$ $\mid $ should provide a direct answer as to the degree of
validity of the principle in question.

\noindent \textit{Remarks on particle deflection experiment}

The \ main concern, which is still far from resolved, is whether realistic
experiments for observing gravitational deflection of massive particles can
be devised or not. Here, we only speculate on some possibilities. The most
important requisite in this context is to generate a beam of suitable test
particles. Charged particles like protons or electrons have to be excluded
as test particles because they suffer electromagnetic interactions by the
interplanetary magnetic field. Among neutral particles, neutrinos are
unlikely to serve the purpose as their speeds are almost, if not exactly,
the same as the speed of light. Thus, neutrons seem to be the only feasible
candidate. They are known to be produced during solar flares but they can at
best be used to study the gravitational deflection by an intermediate
planet. If astrophysical sources of neutrons other than the sun are detected
in future experiments, the problem of searching the test particle beam would
be resolved automatically. Otherwise, one might hope to generate the beam
only artificially. However, since neutrons are unstable with a mean lifetime
of 886 $\sec $, only neutrons with a minimum speed of 0.75$c$ can be used as
test particles so that they do not decay during the travel from one
micro-spacecraft to another. Though in (man-made) accelerator experiments
(at earth) neutrons can be accelerated to such speeds, it seems improbable
at the present stage of technology that neutrons can be accelerated to such
high energies from a micro-spacecraft. This is a challenge for the future.

Alternatively, stable massive objects, such as a bullet, can also be used as
test particles but they must have a minimum speed of $\sim 6\times
10^{7}cm\sec ^{-1}$ so that its total energy remains positive throughout the
path (from one micro-spacecraft/earth to another spacecraft) and would not
be captured by solar gravity. The fastest man-made object (Helios 2 solar
probe) has a speed of about $7\times 10^{6}cm\sec ^{-1}$. However, as
already mentioned, accelerating a material object to speeds of $\sim
10^{8}cm\sec ^{-1}$ and achieving the required level of $\mu$ arcsec
accuracy in the deflection angle is completely beyond present technical
feasibility.

\section{Acknowledgements} The authors would like to thank an anonymous referee for useful comments and suggestions.

\appendix
\section{}

The standard equations for a geodesic, namely 
\begin{equation}
\frac{d^{2}x^{\lambda }}{ds^{2}}+\Gamma _{\mu \nu }^{\lambda }\frac{dx^{\mu }%
}{ds}\frac{dx^{\nu }}{ds}=0
\end{equation}%
for the general metric (1) become 
\begin{equation}
\frac{d^{2}\rho }{ds^{2}}+\frac{A^{^{\prime }}}{2A}\left( \frac{d\rho }{ds}%
\right) ^{2}-\rho \left( 1+\frac{\rho A^{^{\prime }}}{2A}\right) =0
\end{equation}

\begin{equation}
\frac{d^{2}\theta }{ds^{2}}+\left( \frac{2}{\rho }+\frac{A^{^{\prime }}}{A}%
\right) \frac{d\rho }{ds}\frac{d\theta }{ds}-sin\theta cos\theta \left( 
\frac{d\phi }{ds}\right) ^{2}=0
\end{equation}%
\begin{equation}
\frac{d^{2}\phi }{ds^{2}}+\left( \frac{2}{\rho }+\frac{A^{^{\prime }}}{A}%
\right) \frac{d\rho }{ds}\frac{d\theta }{ds}+2cot\theta \frac{d\theta }{ds}%
\frac{d\phi }{ds}=0
\end{equation}%
\begin{equation}
\frac{d^{2}t}{ds^{2}}+\frac{B^{^{\prime }}}{B}\frac{d\rho }{ds}\frac{dt}{ds}%
=0
\end{equation}%
(primes denoting differentiation with respect to $\rho $). If we choose $%
\theta =\pi /2$ and $d\theta /ds=0$ initially, Eq.(A3) warrants that they
would remain the same always. Thus normalizing time coordinate suitably, one
obtains for orbits in the equatorial plane from Eq.(A5) 
\begin{equation}
\frac{dt}{ds}=B^{-1}
\end{equation}%
Integrating Eq.(A4) 
\begin{equation}
\rho ^{2}\frac{d\phi }{ds}=A^{-1}J^{2}
\end{equation}%
where $J$ is a constant of integration. From Eqs.(A2), (A6) and (A7), one
finally obtains 
\begin{equation}
\frac{1}{A\rho ^{4}}\left( \frac{d\rho }{d\phi }\right) ^{2}+\frac{1}{A\rho
^{2}}-\frac{1}{J^{2}}\left( \frac{1}{B}-E\right) =0
\end{equation}%
which leads to Eq.(3). $J$ can be conveniently expressed in terms of
distance at closest approach. At the point of closest approach, $d\rho
/d\phi $ vanishes. Using this in Eq.(A8), one recovers Eq.(4).

\end{document}